\documentclass[traditabstract]{aa}

\usepackage{graphicx}
\usepackage{txfonts}
\usepackage{natbib}
  \bibpunct{(}{)}{;}{a}{}{,}
\usepackage{lscape}

\newcommand{\feh}{\mathrm{[Fe/H]}}
\newcommand{\teff}{T_\mathrm{eff}}
\newcommand{\logg}{\log g}

\newcommand{\fei}{Fe~\textsc{i}}
\newcommand{\feii}{Fe~\textsc{ii}}
\newcommand{\tc}{T_\mathrm{C}}

\begin{document}

\title{Accurate abundance patterns of solar twins and analogs\thanks{Figure 1 and Tables 1--4 are available online at the CDS.}}
\subtitle{Does the anomalous solar chemical composition come from planet formation?}

\titlerunning{Abundance patterns of solar twins}
\authorrunning{Ram\'{\i}rez et al.}

\author{I. Ram\'{\i}rez \inst{1} \and
	J. Mel\'endez   \inst{2} \and
	M. Asplund      \inst{1}
       }

\institute{Max Planck Institute for Astrophysics,
           Postfach 1317, 85741 Garching, Germany;
	   \email{ivan@mpa-garching.mpg.de} 
	   \and
	   Centro de Astrof\'{\i}sica da Universidade do Porto, 
	   Rua das Estrelas 4150-762 Porto, Portugal
          }

\date{Received; accepted}

\abstract
{
We derive the abundance of 19 elements in a sample of 64 stars with fundamental parameters very similar to solar, which minimizes the impact of systematic errors in our spectroscopic 1D-LTE differential analysis, using high-resolution ($R\simeq60,000$), high signal-to-noise ratio ($S/N\simeq200$) spectra. The estimated errors in the elemental abundances relative to solar are as small as $\simeq0.025$\,dex. The abundance ratios [X/Fe] as a function of $\feh$ agree closely with previously established patterns of Galactic thin-disk chemical evolution. Interestingly, the majority of our stars show a significant correlation between [X/Fe] and condensation temperature ($\tc$). In the sample of 22 stars with parameters closest to solar, we find that, on average, low $\tc$ elements are depleted with respect to high $\tc$ elements in the solar twins relative to the Sun by about 0.08\,dex ($\simeq20\,\%$). An increasing trend is observed for the abundances as a function of $\tc$ for $900<\tc<1800$\,K, while abundances of lower $\tc$ elements appear to be roughly constant. We speculate that this is a signature of the planet formation that occurred around the Sun but not in the majority of solar twins. If this hypothesis is correct, stars with planetary systems like ours, although rare (frequency of $\simeq15$\,\%), may be identified through a very detailed inspection of the chemical compositions of their host stars.
}


\maketitle

\section{Introduction}

Standard spectroscopic abundance analyses suffer from a variety of systematic errors that are difficult to remove. Using the highest quality data, errors in stellar parameters, atomic/molecular data, the use of static/homogeneous model atmospheres, and the assumption of local thermodynamic equilibrium set an optimistic lower limit of about 0.05~dex ($\sim10$\,\%) to the accuracy of abundance determinations \citep[e.g.,][]{asplund05:review}. A simple way to minimize the impact of these uncertainties is to perform differential analyses of stars that are very similar to each other so that systematic errors are largely cancelled out. Naturally, attempts have been made using stars with parameters that are very similar to solar, the so-called solar twins \citep[e.g.,][]{melendez06:twins,melendez07:twins}. Recent work on very high-quality spectra of a few solar twins suggests that it is even possible to reach 0.01\,dex accuracy ($\sim2$\,\%, \citealt{melendez09:twins}). This level of precision can be useful for revealing the fine details of abundance trends and, perhaps more importantly, to determine whether the solar chemical composition is anomalous.

The use of the Sun as a reference star is understandable. Its basic properties (effective temperature, luminosity, mass, radius, and age) are very well known, and spectra of high quality are available or can be easily acquired. Defining a good sample of solar twins is a more difficult task \citep{cayrel96}. Stars with fundamental parameters very similar to solar exist \citep[e.g.,][]{porto97,melendez06:twins,takeda07,melendez07:twins}, yet when more detailed analyses of their chemical compositions or evolutionary states are made, some differences arise, such as the apparently low Li abundance of the Sun and the older age of HIP\,56948, the best solar twin known to date \citep{melendez07:twins,takeda09}. However, solar twins with Li abundances and ages similar to those of the Sun most likely exist \citep{pasquini08}, while the age of HIP\,56948 may have been overestimated \citep{donascimento09}. In any case, although the Sun is probably not unique, it does not seem to be a very common object in the solar neighborhood \citep[e.g.,][]{gustafsson08:sun_unique}.

Determining the detailed chemical composition of the Sun compared to other stars in the solar vicinity will help us determine how unusual our star is, and perhaps even why. Previous studies have been inconclusive because of the systematic errors described above, and because the relevant differences may be small. Here we derive precise abundances for a carefully selected sample of solar twins and analogs, for which systematic errors are minimized using differential analysis, and speculate about the nature of the abundance trends found.

\section{Sample, data, and analysis}

Our sample stars were selected from the \textit{Hipparcos} catalog by applying constraints on color, based on the color-$\teff$ calibrations by \cite{ramirez05b}, corrected by suspected zero point errors \citep{casagrande09:irfm}, trigonometric parallaxes, and literature values for $\feh$ and chromospheric activity, if available. About 100 stars satisfied our selection criteria and, given our observational constraints, data for 64 of them were acquired. We also observed the asteroids Vesta and Ceres as solar reference.

Spectra were obtained with the Robert~G.~Tull coud\'e spectrograph \citep{tull95} on the 2.7\,m Harlan J. Smith Telescope at McDonald Observatory in April, October, and November of 2007. The spectral resolution is $R=\lambda/d\lambda\simeq60,000$ and the wavelength coverage 3800--9125\,\AA, with gaps between echelle orders of 10--100\,\AA\ for wavelengths longer than 6100\,\AA. Typically two exposures of 20--30 minutes each were obtained per star with a resulting signal-to-noise ratio ($S/N$) of at least 150 (median $S/N\simeq200$). Equivalent widths ($EW$s) were measured in the spectra (reduced, wavelength calibrated, and normalized using standard techniques) employing four different methods. From comparisons of $EW$ determinations in different spectra for the same star, we estimate that the individual errors in $EW$ are between 2 and 10\,\%, depending on $S/N$ and $EW$.

We compiled a line list of features that are mostly unblended in the spectra of solar-type stars. This line list includes 25 \fei\ lines covering a wide range of excitation potential ($\mathrm{EP}=0$ to 5\,eV) and line strength ($EW=10$ to 120\,m\,\AA), as well as five \feii\ features, which facilitate precise determination of stellar parameters (see below). The majority of the other elements analyzed in this work have between three and eleven available features (C, O, Al, Si, S, Ca, Sc, Ti, V, Cr, Mn, Ni, Ba), while the rest have less than three (Na, Cu, Zn, Y, Zr).

Effective temperatures, surface gravities, and microturbulent velocities were determined with the standard spectroscopic technique of excitation/ionization balance of iron lines. Recently computed Kurucz model atmospheres \citep{castelli03} and the spectrum synthesis program MOOG \citep{sneden73} were used to determine the abundance of iron ($\feh$)\footnote{We adopt: $A_\mathrm{X}=\log n_\mathrm{X}/n_\mathrm{H}$+12, where $n_\mathrm{X}$ is the number density of the element X; $\mathrm{[X/H]}=A_\mathrm{X}-A_\mathrm{X}^\odot$; and $\mathrm{[X/Fe]}=\mathrm{[X/H]}-\mathrm{[Fe/H]}$.} from the \fei\ and \feii\ lines. For a given set of parameters, we measured the $\feh$ vs. excitation potential and $\feh$ vs. reduced equivalent width ($REW=\log EW/\lambda$) slopes for the \fei\ lines, as well as the difference in the mean $\feh$ obtained from the \fei\ and \feii\ lines separately. The stellar parameters were then modified iteratively so that the slopes and the \fei\ minus \feii\ difference approached zero. This procedure was done without subjective human interaction and led to unique solutions, even though our \fei\ linelist showed a mild correlation between EP and \textit{REW}. The 1-$\sigma$ uncertainty of the slopes and \fei\ minus \feii\ difference were used to determine the observational errors (average values given here): $\sigma(\teff)=50$\,K, $\sigma(\logg)=0.07$\,dex, and $\sigma(\feh)=0.024$\,dex. Hereafter, the 22 stars with $\teff$ within 100\,K, $\logg$ within 0.1\,dex and $\feh$ within 0.1\,dex of the solar values are referred to as solar twins.

The abundances of the 18 other elements were calculated with MOOG, using the appropriate Kurucz model atmosphere and the curve-of-growth technique. The solar abundances, derived from our asteroid spectra, were used to determine relative abundances of each star on a line-by-line basis. We did not require extremely accurate absolute solar abundances for our analysis, which is entirely differential. Since almost all lines analyzed by us are well within the linear part of the curve of growth (the exceptions being a few strong \fei\ and Ba~\textsc{ii} lines), the uncertainties in our derived solar abundances and/or adopted transition probabilities are irrelevant. The average and standard deviation of the line-by-line relative abundances were adopted as the final abundance, [X/H], and error. On average, the [X/H] values have line-by-line uncertainties of $\simeq0.025$\,dex. Errors in the parameters affect [X/H] and $\feh$ similarly so that [X/Fe] abundance ratios are relatively insensitive to those uncertainties. The average error in [X/Fe] is $\simeq0.03$\,dex, which is dominated by the line-by-line scatter of the X and Fe elemental abundances. To improve the accuracy, non-LTE corrections were applied to the derived oxygen abundances using the results by \cite{ramirez07}, while hyperfine structure was taken into account in the synthesis of Mn and V lines \citep{prochaska00,johnson06}.

Our derived stellar parameters and abundances are available online (Tables~\ref{t:parameters} to \ref{t:abundances3}).

\onltab{1}{
\begin{table*}
\centering \tiny
\caption{Derived stellar parameters, average abundance ratios [X/Fe] of volatile elements ($\tc<900$\,K), and slope of the [X/Fe] vs. $\tc$ relation for refractory elements ($\tc>900$\,K). Errors in $\feh$ correspond to the 1-$\sigma$ scatter of the line-by-line abundances.}
\begin{tabular}{rcrrrr} \hline\hline
HIP & $\teff$ & $\logg$ & \multicolumn{1}{c}{$\feh$} & $\tc<900$\,K                     & $\tc>900$\,K \\
    &   (K)   &  [cgs]  &                            & \multicolumn{1}{c}{avg. [X/Fe]}  & \multicolumn{1}{c}{$\tc$ slope} \\ \hline
   348 & 5777 & 4.41 & -0.132$\pm$0.024 &  0.041$\pm$0.025 &  0.097$\pm$0.024 \\
   996 & 5860 & 4.38 & -0.001$\pm$0.022 &  0.024$\pm$0.065 &  0.018$\pm$0.033 \\
  1499 & 5751 & 4.33 &  0.180$\pm$0.043 & -0.008$\pm$0.043 & -0.138$\pm$0.038 \\
  2131 & 5720 & 4.38 & -0.209$\pm$0.026 &  0.087$\pm$0.025 &  0.160$\pm$0.030 \\
  2894 & 5820 & 4.54 & -0.023$\pm$0.025 & -0.034$\pm$0.025 &  0.127$\pm$0.028 \\
  4909 & 5836 & 4.44 &  0.018$\pm$0.024 & -0.075$\pm$0.033 &  0.155$\pm$0.030 \\
  5134 & 5779 & 4.49 & -0.187$\pm$0.023 & -0.006$\pm$0.033 &  0.175$\pm$0.022 \\
  6407 & 5787 & 4.47 & -0.084$\pm$0.011 & -0.017$\pm$0.025 &  0.127$\pm$0.025 \\
  7245 & 5843 & 4.53 &  0.105$\pm$0.023 & -0.012$\pm$0.037 &  0.071$\pm$0.030 \\
  8507 & 5720 & 4.44 & -0.079$\pm$0.026 & -0.049$\pm$0.026 &  0.196$\pm$0.026 \\
  8841 & 5676 & 4.50 & -0.127$\pm$0.021 &  0.046$\pm$0.038 &  0.161$\pm$0.024 \\
  9349 & 5826 & 4.50 &  0.010$\pm$0.019 & -0.068$\pm$0.025 &  0.132$\pm$0.020 \\
 10710 & 5817 & 4.39 & -0.130$\pm$0.022 &  0.037$\pm$0.025 &  0.090$\pm$0.024 \\
 11728 & 5747 & 4.37 &  0.049$\pm$0.022 & -0.014$\pm$0.029 &  0.020$\pm$0.032 \\
 11915 & 5793 & 4.45 & -0.052$\pm$0.021 & -0.040$\pm$0.025 &  0.075$\pm$0.025 \\
 14614 & 5794 & 4.42 & -0.103$\pm$0.022 &  0.016$\pm$0.030 &  0.092$\pm$0.034 \\
 14632 & 6019 & 4.25 &  0.124$\pm$0.024 & -0.005$\pm$0.034 &  0.006$\pm$0.033 \\
 18261 & 5897 & 4.45 &  0.001$\pm$0.017 & -0.040$\pm$0.069 &  0.099$\pm$0.027 \\
 22528 & 5683 & 4.33 & -0.351$\pm$0.035 &  0.205$\pm$0.039 &  0.112$\pm$0.036 \\
 23835 & 5736 & 4.14 & -0.183$\pm$0.020 &  0.107$\pm$0.025 &  0.228$\pm$0.030 \\
 25670 & 5757 & 4.37 &  0.066$\pm$0.018 & -0.037$\pm$0.039 &  0.055$\pm$0.022 \\
 28336 & 5713 & 4.53 & -0.175$\pm$0.027 & -0.024$\pm$0.048 &  0.088$\pm$0.024 \\
 38072 & 5839 & 4.53 &  0.059$\pm$0.037 & -0.039$\pm$0.044 &  0.087$\pm$0.036 \\
 38228 & 5688 & 4.52 & -0.003$\pm$0.026 & -0.132$\pm$0.025 &  0.096$\pm$0.030 \\
 42438 & 5889 & 4.47 & -0.036$\pm$0.029 & -0.107$\pm$0.062 &  0.104$\pm$0.034 \\
 44324 & 5937 & 4.51 & -0.014$\pm$0.021 & -0.087$\pm$0.056 &  0.193$\pm$0.050 \\
 46066 & 5709 & 4.49 & -0.073$\pm$0.039 & -0.024$\pm$0.026 &  0.163$\pm$0.031 \\
 49572 & 5831 & 4.33 &  0.008$\pm$0.021 &  0.010$\pm$0.048 &  0.037$\pm$0.038 \\
 49756 & 5900 & 4.60 &  0.081$\pm$0.037 & -0.056$\pm$0.029 &  0.077$\pm$0.028 \\
 52040 & 5785 & 4.51 & -0.090$\pm$0.021 &  0.013$\pm$0.054 &  0.157$\pm$0.024 \\
 52137 & 5842 & 4.56 &  0.069$\pm$0.026 & -0.108$\pm$0.038 &  0.205$\pm$0.025 \\
 55459 & 5835 & 4.39 &  0.030$\pm$0.022 & -0.044$\pm$0.032 &  0.085$\pm$0.029 \\
 56948 & 5837 & 4.47 &  0.044$\pm$0.027 & -0.032$\pm$0.025 & -0.003$\pm$0.024 \\
 56997 & 5575 & 4.55 & -0.026$\pm$0.030 & -0.011$\pm$0.055 &  0.135$\pm$0.030 \\
 60314 & 5874 & 4.52 &  0.115$\pm$0.033 & -0.127$\pm$0.029 &  0.132$\pm$0.025 \\
 62175 & 5854 & 4.44 &  0.144$\pm$0.021 & -0.069$\pm$0.051 &  0.085$\pm$0.020 \\
 64150 & 5748 & 4.34 &  0.050$\pm$0.020 & -0.038$\pm$0.083 &  0.014$\pm$0.025 \\
 64497 & 5860 & 4.56 &  0.117$\pm$0.037 & -0.079$\pm$0.066 &  0.007$\pm$0.031 \\
 64794 & 5743 & 4.33 & -0.105$\pm$0.027 &  0.092$\pm$0.025 &  0.050$\pm$0.027 \\
 72659 & 5494 & 4.52 & -0.149$\pm$0.039 &  0.042$\pm$0.097 &  0.194$\pm$0.036 \\
 73815 & 5799 & 4.31 &  0.021$\pm$0.022 &  0.016$\pm$0.028 &  0.066$\pm$0.026 \\
 74341 & 5853 & 4.51 &  0.084$\pm$0.026 & -0.064$\pm$0.044 &  0.069$\pm$0.028 \\
 77466 & 5700 & 4.40 & -0.280$\pm$0.028 &  0.214$\pm$0.039 &  0.082$\pm$0.028 \\
 78028 & 5879 & 4.57 & -0.035$\pm$0.041 &  0.005$\pm$0.025 &  0.079$\pm$0.037 \\
 78680 & 5923 & 4.57 & -0.004$\pm$0.027 & -0.059$\pm$0.027 &  0.130$\pm$0.028 \\
 79186 & 5709 & 4.27 & -0.119$\pm$0.024 &  0.108$\pm$0.089 &  0.117$\pm$0.029 \\
 79672 & 5848 & 4.46 &  0.055$\pm$0.019 & -0.047$\pm$0.051 &  0.074$\pm$0.029 \\
 81512 & 5790 & 4.46 & -0.019$\pm$0.025 & -0.010$\pm$0.047 &  0.111$\pm$0.028 \\
 83601 & 6090 & 4.40 &  0.051$\pm$0.031 & -0.098$\pm$0.049 &  0.106$\pm$0.024 \\
 88194 & 5746 & 4.38 & -0.067$\pm$0.015 & -0.037$\pm$0.037 &  0.071$\pm$0.021 \\
 88427 & 5810 & 4.42 & -0.160$\pm$0.025 &  0.118$\pm$0.032 &  0.106$\pm$0.026 \\
 89443 & 5796 & 4.48 & -0.026$\pm$0.038 & -0.055$\pm$0.025 &  0.113$\pm$0.018 \\
 96895 & 5825 & 4.33 &  0.096$\pm$0.026 &  0.009$\pm$0.025 &  0.013$\pm$0.024 \\
 96901 & 5750 & 4.34 &  0.052$\pm$0.021 &  0.003$\pm$0.025 & -0.069$\pm$0.023 \\
100963 & 5815 & 4.49 &  0.018$\pm$0.019 & -0.052$\pm$0.025 &  0.099$\pm$0.032 \\
102152 & 5746 & 4.40 & -0.013$\pm$0.031 &  0.018$\pm$0.041 &  0.052$\pm$0.026 \\
104504 & 5836 & 4.50 & -0.154$\pm$0.022 & -0.022$\pm$0.040 &  0.170$\pm$0.027 \\
107350 & 6034 & 4.48 & -0.016$\pm$0.024 & -0.103$\pm$0.037 &  0.182$\pm$0.038 \\
108708 & 5875 & 4.51 &  0.152$\pm$0.024 & -0.108$\pm$0.032 &  0.138$\pm$0.035 \\
108996 & 5838 & 4.50 &  0.056$\pm$0.027 & -0.049$\pm$0.061 &  0.210$\pm$0.032 \\
109931 & 5739 & 4.29 &  0.037$\pm$0.026 &  0.028$\pm$0.033 & -0.064$\pm$0.029 \\
113357 & 5832 & 4.38 &  0.215$\pm$0.023 & -0.034$\pm$0.049 & -0.044$\pm$0.025 \\
115604 & 5821 & 4.43 &  0.137$\pm$0.019 &  0.008$\pm$0.047 & -0.056$\pm$0.030 \\
118159 & 5905 & 4.55 & -0.010$\pm$0.022 & -0.082$\pm$0.038 &  0.313$\pm$0.030 \\

\hline
\end{tabular}
\label{t:parameters}
\end{table*}
}

\onltab{2}{
\begin{table*}
\centering \tiny
\caption{Abundance ratios [X/Fe] for C, O, Na, Al, Si, and S. Errors correspond to the 1-$\sigma$ scatter of the line-by-line abundances. A conservative value of 0.050 for the error was assumed when only one line was available.}
\begin{tabular}{rrrrrrr} \hline\hline
HIP & \multicolumn{1}{c}{C} & \multicolumn{1}{c}{O} & \multicolumn{1}{c}{Na} & \multicolumn{1}{c}{Al} & \multicolumn{1}{c}{Si} & \multicolumn{1}{c}{S} \\ \hline
   348& 0.057$\pm$0.037& 0.050$\pm$0.045&-0.039$\pm$0.021& 0.032$\pm$0.024&-0.005$\pm$0.042& 0.010$\pm$0.027\\
   996& 0.020$\pm$0.031&-0.011$\pm$0.012& 0.008$\pm$0.025& 0.060$\pm$0.022& 0.029$\pm$0.030&-0.030$\pm$0.026\\
  1499&-0.012$\pm$0.030&-0.068$\pm$0.036& 0.118$\pm$0.025& 0.066$\pm$0.043& 0.015$\pm$0.035& 0.019$\pm$0.026\\
  2131& 0.102$\pm$0.030& 0.094$\pm$0.032&-0.060$\pm$0.025& 0.058$\pm$0.029& 0.020$\pm$0.034& 0.056$\pm$0.070\\
  2894&-0.044$\pm$0.038&-0.057$\pm$0.035&-0.078$\pm$0.016&-0.059$\pm$0.042&-0.042$\pm$0.026&-0.026$\pm$0.025\\
  4909&-0.122$\pm$0.013&-0.045$\pm$0.035&-0.113$\pm$0.021&-0.087$\pm$0.035&-0.037$\pm$0.037&-0.069$\pm$0.052\\
  5134&-0.042$\pm$0.050& 0.022$\pm$0.038&-0.089$\pm$0.012&-0.049$\pm$0.026&-0.015$\pm$0.040& 0.022$\pm$0.050\\
  6407&-0.043$\pm$0.024&-0.001$\pm$0.031&-0.045$\pm$0.014&-0.008$\pm$0.027& 0.004$\pm$0.031&-0.008$\pm$0.032\\
  7245&-0.044$\pm$0.013&-0.042$\pm$0.043& 0.005$\pm$0.025&-0.034$\pm$0.029&-0.026$\pm$0.049& 0.013$\pm$0.021\\
  8507&-0.082$\pm$0.050&-0.019$\pm$0.059&-0.084$\pm$0.016&-0.007$\pm$0.034& 0.004$\pm$0.079&-0.048$\pm$0.053\\
  8841&-0.000$\pm$0.025& 0.032$\pm$0.081&-0.086$\pm$0.025&-0.004$\pm$0.032&-0.037$\pm$0.043& 0.082$\pm$0.028\\
  9349&-0.086$\pm$0.025&-0.091$\pm$0.118&-0.061$\pm$0.011&-0.078$\pm$0.015&-0.046$\pm$0.046&-0.051$\pm$0.061\\
 10710& 0.051$\pm$0.024& 0.041$\pm$0.030&-0.041$\pm$0.025&-0.034$\pm$0.025&-0.009$\pm$0.029& 0.049$\pm$0.063\\
 11728&-0.047$\pm$0.017&-0.030$\pm$0.025&-0.024$\pm$0.025& 0.008$\pm$0.028&-0.002$\pm$0.038& 0.011$\pm$0.025\\
 11915&-0.046$\pm$0.012&-0.032$\pm$0.061&-0.062$\pm$0.018&-0.057$\pm$0.015&-0.022$\pm$0.031&-0.013$\pm$0.064\\
 14614& 0.007$\pm$0.046&-0.012$\pm$0.046&-0.047$\pm$0.029& 0.014$\pm$0.059&-0.006$\pm$0.044& 0.011$\pm$0.025\\
 14632&-0.003$\pm$0.041&-0.034$\pm$0.031& 0.042$\pm$0.025& 0.003$\pm$0.038&-0.010$\pm$0.044&-0.026$\pm$0.078\\
 18261&-0.098$\pm$0.048&-0.040$\pm$0.036&-0.089$\pm$0.025&-0.066$\pm$0.013&-0.024$\pm$0.025&-0.079$\pm$0.086\\
 22528& 0.253$\pm$0.160& 0.192$\pm$0.052&-0.037$\pm$0.025& 0.162$\pm$0.041& 0.063$\pm$0.036& 0.160$\pm$0.010\\
 23835& 0.111$\pm$0.052& 0.139$\pm$0.038&-0.017$\pm$0.025& 0.132$\pm$0.021& 0.072$\pm$0.029& 0.090$\pm$0.011\\
 25670&-0.092$\pm$0.025&-0.038$\pm$0.036&-0.041$\pm$0.016&-0.002$\pm$0.053&-0.012$\pm$0.045&-0.012$\pm$0.011\\
 28336&-0.021$\pm$0.025&-0.007$\pm$0.050&-0.051$\pm$0.021&-0.055$\pm$0.024&-0.024$\pm$0.045& 0.023$\pm$0.050\\
 38072&-0.044$\pm$0.035& 0.001$\pm$0.080&-0.038$\pm$0.025&-0.001$\pm$0.046&-0.001$\pm$0.036&-0.013$\pm$0.025\\
 38228&-0.137$\pm$0.050&-0.113$\pm$0.050&-0.065$\pm$0.025&-0.023$\pm$0.020&-0.009$\pm$0.061&\\
 42438&-0.035$\pm$0.025&-0.077$\pm$0.085&-0.081$\pm$0.025&-0.083$\pm$0.037&-0.034$\pm$0.047&-0.150$\pm$0.108\\
 44324&-0.098$\pm$0.020&-0.136$\pm$0.017&&-0.156$\pm$0.076&-0.119$\pm$0.064&\\
 46066&-0.015$\pm$0.040&-0.025$\pm$0.035&-0.059$\pm$0.025&-0.049$\pm$0.071&-0.030$\pm$0.027&-0.058$\pm$0.104\\
 49572& 0.029$\pm$0.018&-0.048$\pm$0.025& 0.020$\pm$0.025& 0.040$\pm$0.034& 0.008$\pm$0.048&-0.006$\pm$0.020\\
 49756&-0.020$\pm$0.033&-0.066$\pm$0.036&-0.029$\pm$0.025&-0.037$\pm$0.045&-0.031$\pm$0.029&-0.051$\pm$0.011\\
 52040& 0.088$\pm$0.099& 0.003$\pm$0.043&-0.079$\pm$0.012&-0.056$\pm$0.052&-0.024$\pm$0.039&-0.039$\pm$0.050\\
 52137&-0.157$\pm$0.024&-0.065$\pm$0.024&-0.086$\pm$0.020&-0.064$\pm$0.025&-0.023$\pm$0.044&-0.106$\pm$0.046\\
 55459&-0.065$\pm$0.019&-0.015$\pm$0.052&-0.036$\pm$0.025& 0.023$\pm$0.075&-0.015$\pm$0.033&-0.078$\pm$0.013\\
 56948&-0.025$\pm$0.039&-0.033$\pm$0.039& 0.008$\pm$0.015&-0.010$\pm$0.058&-0.027$\pm$0.034&-0.053$\pm$0.014\\
 56997&-0.033$\pm$0.013& 0.023$\pm$0.025&-0.052$\pm$0.025&-0.103$\pm$0.048&-0.041$\pm$0.026& 0.042$\pm$0.041\\
 60314&-0.149$\pm$0.023&-0.093$\pm$0.040&-0.057$\pm$0.018&-0.028$\pm$0.033&-0.014$\pm$0.028&-0.112$\pm$0.058\\
 62175&-0.111$\pm$0.014&-0.108$\pm$0.025&-0.032$\pm$0.011&-0.013$\pm$0.032&-0.014$\pm$0.040&-0.054$\pm$0.032\\
 64150& 0.024$\pm$0.025&-0.151$\pm$0.054& 0.004$\pm$0.025& 0.008$\pm$0.022&-0.005$\pm$0.040& 0.024$\pm$0.013\\
 64497&-0.118$\pm$0.012& 0.002$\pm$0.025&-0.088$\pm$0.025&-0.048$\pm$0.050&-0.026$\pm$0.032&-0.056$\pm$0.057\\
 64794& 0.101$\pm$0.023& 0.097$\pm$0.011&-0.042$\pm$0.025& 0.021$\pm$0.020& 0.019$\pm$0.038& 0.069$\pm$0.025\\
 72659&-0.008$\pm$0.030& 0.153$\pm$0.031&-0.079$\pm$0.025&-0.062$\pm$0.028&-0.027$\pm$0.024& 0.086$\pm$0.062\\
 73815&-0.021$\pm$0.045& 0.020$\pm$0.020&-0.028$\pm$0.025& 0.020$\pm$0.023& 0.013$\pm$0.030& 0.020$\pm$0.041\\
 74341&-0.059$\pm$0.056&-0.054$\pm$0.038&-0.024$\pm$0.023&-0.049$\pm$0.041&-0.029$\pm$0.030&-0.018$\pm$0.028\\
 77466& 0.250$\pm$0.138& 0.220$\pm$0.030& 0.016$\pm$0.025& 0.190$\pm$0.055& 0.095$\pm$0.025& 0.158$\pm$0.100\\
 78028&-0.013$\pm$0.024& 0.019$\pm$0.038&-0.018$\pm$0.025&-0.011$\pm$0.039& 0.022$\pm$0.064& 0.026$\pm$0.017\\
 78680&-0.089$\pm$0.049&-0.035$\pm$0.023&-0.115$\pm$0.026&-0.116$\pm$0.031&-0.057$\pm$0.041&-0.075$\pm$0.064\\
 79186& 0.154$\pm$0.166& 0.190$\pm$0.040&-0.046$\pm$0.025& 0.108$\pm$0.019& 0.040$\pm$0.032&-0.014$\pm$0.024\\
 79672&-0.113$\pm$0.033&-0.041$\pm$0.025&-0.042$\pm$0.025&-0.038$\pm$0.039&-0.010$\pm$0.031&-0.047$\pm$0.027\\
 81512&-0.025$\pm$0.023&-0.024$\pm$0.018&-0.048$\pm$0.025& 0.018$\pm$0.029&-0.026$\pm$0.031&-0.048$\pm$0.040\\
 83601&-0.073$\pm$0.068&-0.051$\pm$0.030&-0.085$\pm$0.014&-0.064$\pm$0.018&-0.040$\pm$0.037&-0.102$\pm$0.113\\
 88194&-0.050$\pm$0.063& 0.018$\pm$0.011&-0.044$\pm$0.025& 0.023$\pm$0.039&-0.003$\pm$0.026&-0.060$\pm$0.029\\
 88427& 0.125$\pm$0.103& 0.136$\pm$0.024& 0.005$\pm$0.025& 0.144$\pm$0.036& 0.045$\pm$0.031& 0.071$\pm$0.019\\
 89443&-0.068$\pm$0.018&-0.059$\pm$0.025&-0.040$\pm$0.013& 0.088$\pm$0.019& 0.004$\pm$0.012&-0.072$\pm$0.038\\
 96895& 0.007$\pm$0.052&-0.002$\pm$0.025& 0.021$\pm$0.025& 0.068$\pm$0.020& 0.038$\pm$0.046&-0.002$\pm$0.040\\
 96901& 0.000$\pm$0.075& 0.005$\pm$0.021& 0.043$\pm$0.037& 0.072$\pm$0.019& 0.019$\pm$0.064& 0.010$\pm$0.016\\
100963&-0.080$\pm$0.028&-0.058$\pm$0.043&-0.070$\pm$0.025&-0.050$\pm$0.055&-0.049$\pm$0.036&-0.022$\pm$0.028\\
102152&-0.004$\pm$0.057&-0.021$\pm$0.032&-0.034$\pm$0.029& 0.032$\pm$0.016& 0.016$\pm$0.055& 0.022$\pm$0.034\\
104504&-0.070$\pm$0.040& 0.028$\pm$0.041&-0.082$\pm$0.023&-0.076$\pm$0.033&-0.035$\pm$0.031&-0.026$\pm$0.058\\
107350&-0.086$\pm$0.032&-0.060$\pm$0.043&-0.091$\pm$0.025&-0.042$\pm$0.090&-0.027$\pm$0.045&-0.125$\pm$0.024\\
108708&-0.142$\pm$0.030&-0.124$\pm$0.049&-0.077$\pm$0.025&-0.071$\pm$0.020&-0.035$\pm$0.049&-0.069$\pm$0.051\\
108996&-0.063$\pm$0.012& 0.024$\pm$0.035&-0.083$\pm$0.035&-0.037$\pm$0.050&-0.035$\pm$0.043&-0.035$\pm$0.073\\
109931&-0.015$\pm$0.042& 0.029$\pm$0.033& 0.039$\pm$0.023& 0.071$\pm$0.016& 0.057$\pm$0.045& 0.064$\pm$0.044\\
113357&-0.057$\pm$0.020&-0.093$\pm$0.028& 0.065$\pm$0.025& 0.040$\pm$0.035& 0.013$\pm$0.030& 0.008$\pm$0.027\\
115604& 0.012$\pm$0.058&-0.043$\pm$0.040& 0.030$\pm$0.030& 0.041$\pm$0.032& 0.036$\pm$0.041&-0.006$\pm$0.016\\
118159&-0.091$\pm$0.044&-0.045$\pm$0.040&-0.158$\pm$0.025&-0.104$\pm$0.026&-0.062$\pm$0.039&-0.061$\pm$0.067\\

\hline
\end{tabular}
\label{t:abundances1}
\end{table*}
}

\onltab{3}{
\begin{table*}
\centering \tiny
\caption{Abundance ratios [X/Fe] for Ca, Sc, Ti, V, Cr, and Mn. Errors correspond to the 1-$\sigma$ scatter of the line-by-line abundances. A conservative value of 0.050 for the error was assumed when only one line was available.}
\begin{tabular}{rrrrrrr} \hline\hline
HIP & \multicolumn{1}{c}{Ca} & \multicolumn{1}{c}{Sc} & \multicolumn{1}{c}{Ti} & \multicolumn{1}{c}{V} & \multicolumn{1}{c}{Cr} & \multicolumn{1}{c}{Mn} \\ \hline
   348& 0.030$\pm$0.070& 0.036$\pm$0.015& 0.015$\pm$0.041& 0.028$\pm$0.046&-0.011$\pm$0.023&-0.074$\pm$0.017\\
   996& 0.018$\pm$0.037& 0.057$\pm$0.072& 0.018$\pm$0.025& 0.023$\pm$0.013&-0.009$\pm$0.035& 0.002$\pm$0.026\\
  1499& 0.004$\pm$0.033&-0.033$\pm$0.025&-0.022$\pm$0.025&-0.019$\pm$0.020&-0.025$\pm$0.041& 0.062$\pm$0.050\\
  2131& 0.059$\pm$0.040& 0.058$\pm$0.056& 0.064$\pm$0.033& 0.018$\pm$0.041&-0.022$\pm$0.032&-0.072$\pm$0.018\\
  2894& 0.012$\pm$0.020& 0.080$\pm$0.051& 0.010$\pm$0.029& 0.002$\pm$0.019& 0.013$\pm$0.036&-0.033$\pm$0.025\\
  4909& 0.045$\pm$0.052& 0.074$\pm$0.049&-0.028$\pm$0.050&-0.009$\pm$0.038& 0.005$\pm$0.035&-0.028$\pm$0.024\\
  5134& 0.038$\pm$0.029& 0.014$\pm$0.023& 0.020$\pm$0.051&-0.010$\pm$0.039&-0.032$\pm$0.027&-0.096$\pm$0.031\\
  6407& 0.036$\pm$0.019& 0.024$\pm$0.031& 0.032$\pm$0.036& 0.046$\pm$0.038& 0.029$\pm$0.021&-0.028$\pm$0.025\\
  7245& 0.013$\pm$0.045& 0.026$\pm$0.065&-0.001$\pm$0.048& 0.051$\pm$0.040&-0.004$\pm$0.039& 0.030$\pm$0.025\\
  8507& 0.048$\pm$0.015&-0.000$\pm$0.057& 0.038$\pm$0.038& 0.002$\pm$0.021&-0.002$\pm$0.030&-0.075$\pm$0.026\\
  8841& 0.023$\pm$0.069& 0.017$\pm$0.057& 0.016$\pm$0.027& 0.003$\pm$0.033&-0.018$\pm$0.033&-0.062$\pm$0.037\\
  9349& 0.013$\pm$0.038& 0.025$\pm$0.060&-0.002$\pm$0.034& 0.017$\pm$0.013&-0.001$\pm$0.050&-0.021$\pm$0.018\\
 10710& 0.024$\pm$0.039& 0.013$\pm$0.072&-0.027$\pm$0.051& 0.009$\pm$0.043& 0.005$\pm$0.036&-0.046$\pm$0.016\\
 11728& 0.042$\pm$0.027&-0.025$\pm$0.078&-0.014$\pm$0.017& 0.001$\pm$0.030&-0.004$\pm$0.036& 0.005$\pm$0.019\\
 11915& 0.021$\pm$0.031& 0.004$\pm$0.041&-0.015$\pm$0.029& 0.006$\pm$0.030& 0.006$\pm$0.019&-0.033$\pm$0.018\\
 14614& 0.051$\pm$0.039& 0.025$\pm$0.026& 0.006$\pm$0.034&-0.018$\pm$0.019&-0.019$\pm$0.027&-0.037$\pm$0.032\\
 14632&-0.003$\pm$0.079& 0.010$\pm$0.033&-0.052$\pm$0.054&-0.002$\pm$0.038&-0.030$\pm$0.021&-0.000$\pm$0.025\\
 18261& 0.018$\pm$0.028&-0.000$\pm$0.057& 0.002$\pm$0.036&-0.028$\pm$0.015& 0.004$\pm$0.034&-0.036$\pm$0.019\\
 22528& 0.102$\pm$0.045& 0.097$\pm$0.035& 0.134$\pm$0.065& 0.066$\pm$0.046& 0.001$\pm$0.042&-0.146$\pm$0.025\\
 23835& 0.083$\pm$0.028& 0.111$\pm$0.033& 0.102$\pm$0.029& 0.052$\pm$0.027&-0.007$\pm$0.035&-0.132$\pm$0.018\\
 25670& 0.035$\pm$0.024&-0.037$\pm$0.072&-0.011$\pm$0.023&-0.019$\pm$0.024&-0.008$\pm$0.039& 0.004$\pm$0.013\\
 28336& 0.030$\pm$0.061& 0.017$\pm$0.052&-0.001$\pm$0.062& 0.041$\pm$0.040&-0.008$\pm$0.013&-0.074$\pm$0.017\\
 38072& 0.037$\pm$0.091& 0.025$\pm$0.064& 0.027$\pm$0.029& 0.023$\pm$0.038& 0.029$\pm$0.027&-0.010$\pm$0.025\\
 38228& 0.099$\pm$0.063& 0.036$\pm$0.013& 0.037$\pm$0.054& 0.035$\pm$0.027& 0.049$\pm$0.051&-0.012$\pm$0.023\\
 42438& 0.059$\pm$0.041&& 0.026$\pm$0.032& 0.013$\pm$0.083& 0.068$\pm$0.059&-0.081$\pm$0.036\\
 44324&-0.014$\pm$0.028& 0.010$\pm$0.067& 0.029$\pm$0.080& 0.023$\pm$0.050& 0.023$\pm$0.063&-0.083$\pm$0.019\\
 46066& 0.036$\pm$0.027& 0.042$\pm$0.075& 0.066$\pm$0.021& 0.033$\pm$0.028& 0.026$\pm$0.040& 0.000$\pm$0.013\\
 49572& 0.023$\pm$0.036& 0.020$\pm$0.107& 0.015$\pm$0.037& 0.024$\pm$0.034&-0.007$\pm$0.050&-0.011$\pm$0.028\\
 49756& 0.006$\pm$0.061& 0.043$\pm$0.034& 0.038$\pm$0.014& 0.047$\pm$0.025& 0.019$\pm$0.046& 0.003$\pm$0.025\\
 52040& 0.054$\pm$0.097& 0.032$\pm$0.034& 0.043$\pm$0.020& 0.001$\pm$0.033& 0.028$\pm$0.044&-0.046$\pm$0.017\\
 52137& 0.018$\pm$0.023& 0.014$\pm$0.063& 0.042$\pm$0.024& 0.011$\pm$0.025& 0.032$\pm$0.028&-0.029$\pm$0.015\\
 55459& 0.009$\pm$0.032& 0.016$\pm$0.032& 0.028$\pm$0.022& 0.006$\pm$0.017& 0.017$\pm$0.024&-0.029$\pm$0.025\\
 56948&-0.006$\pm$0.034& 0.013$\pm$0.031& 0.023$\pm$0.028&-0.006$\pm$0.031& 0.034$\pm$0.047&-0.005$\pm$0.020\\
 56997& 0.017$\pm$0.020&-0.025$\pm$0.053& 0.032$\pm$0.014& 0.020$\pm$0.012& 0.059$\pm$0.028&-0.002$\pm$0.025\\
 60314& 0.017$\pm$0.036&-0.050$\pm$0.050& 0.070$\pm$0.027& 0.020$\pm$0.021& 0.022$\pm$0.058&-0.013$\pm$0.027\\
 62175& 0.018$\pm$0.020& 0.014$\pm$0.020& 0.042$\pm$0.020& 0.023$\pm$0.025& 0.030$\pm$0.030& 0.012$\pm$0.027\\
 64150& 0.023$\pm$0.035&-0.001$\pm$0.044& 0.013$\pm$0.013&-0.011$\pm$0.022& 0.029$\pm$0.034&-0.004$\pm$0.022\\
 64497& 0.020$\pm$0.033& 0.024$\pm$0.050& 0.003$\pm$0.035& 0.004$\pm$0.015& 0.025$\pm$0.031& 0.002$\pm$0.022\\
 64794& 0.015$\pm$0.045& 0.009$\pm$0.058& 0.023$\pm$0.042&-0.031$\pm$0.024& 0.027$\pm$0.050&-0.047$\pm$0.013\\
 72659& 0.074$\pm$0.029&-0.023$\pm$0.046& 0.042$\pm$0.024& 0.019$\pm$0.019& 0.047$\pm$0.034&-0.036$\pm$0.021\\
 73815& 0.008$\pm$0.019& 0.023$\pm$0.018& 0.036$\pm$0.024&-0.005$\pm$0.023& 0.025$\pm$0.034&-0.025$\pm$0.019\\
 74341&-0.004$\pm$0.023& 0.024$\pm$0.072& 0.027$\pm$0.023& 0.034$\pm$0.019& 0.041$\pm$0.022& 0.026$\pm$0.028\\
 77466& 0.106$\pm$0.037& 0.074$\pm$0.025& 0.195$\pm$0.019& 0.110$\pm$0.035& 0.021$\pm$0.045&-0.136$\pm$0.023\\
 78028& 0.026$\pm$0.069& 0.020$\pm$0.074& 0.042$\pm$0.045& 0.031$\pm$0.055&-0.013$\pm$0.052&-0.015$\pm$0.025\\
 78680& 0.007$\pm$0.031& 0.011$\pm$0.012&-0.005$\pm$0.036&-0.022$\pm$0.035& 0.007$\pm$0.035&-0.054$\pm$0.025\\
 79186& 0.071$\pm$0.015& 0.038$\pm$0.078& 0.100$\pm$0.015& 0.003$\pm$0.033&-0.001$\pm$0.025&-0.086$\pm$0.034\\
 79672&-0.003$\pm$0.024& 0.009$\pm$0.015& 0.031$\pm$0.025& 0.014$\pm$0.012& 0.030$\pm$0.033&-0.009$\pm$0.025\\
 81512& 0.016$\pm$0.033&-0.008$\pm$0.079& 0.042$\pm$0.014&-0.001$\pm$0.024& 0.022$\pm$0.036&-0.054$\pm$0.025\\
 83601& 0.039$\pm$0.034& 0.000$\pm$0.022&-0.019$\pm$0.039&-0.007$\pm$0.053&-0.003$\pm$0.023&-0.052$\pm$0.040\\
 88194& 0.023$\pm$0.023& 0.006$\pm$0.027& 0.043$\pm$0.014&-0.003$\pm$0.027& 0.010$\pm$0.038&-0.042$\pm$0.025\\
 88427& 0.065$\pm$0.033& 0.079$\pm$0.019& 0.135$\pm$0.024& 0.055$\pm$0.027&-0.020$\pm$0.025&-0.124$\pm$0.017\\
 89443& 0.026$\pm$0.033& 0.036$\pm$0.012& 0.102$\pm$0.021& 0.057$\pm$0.028& 0.053$\pm$0.060&-0.039$\pm$0.013\\
 96895& 0.013$\pm$0.036& 0.030$\pm$0.054& 0.029$\pm$0.027& 0.042$\pm$0.058&-0.003$\pm$0.020& 0.034$\pm$0.016\\
 96901& 0.032$\pm$0.045& 0.002$\pm$0.031& 0.021$\pm$0.028& 0.010$\pm$0.024&-0.000$\pm$0.020& 0.002$\pm$0.022\\
100963&-0.009$\pm$0.034&-0.002$\pm$0.032&-0.021$\pm$0.033&-0.018$\pm$0.035&-0.013$\pm$0.023&-0.038$\pm$0.025\\
102152& 0.014$\pm$0.039& 0.017$\pm$0.027& 0.029$\pm$0.031& 0.018$\pm$0.023& 0.010$\pm$0.019&-0.025$\pm$0.013\\
104504& 0.035$\pm$0.030&& 0.011$\pm$0.052& 0.006$\pm$0.015& 0.045$\pm$0.040&-0.093$\pm$0.015\\
107350& 0.045$\pm$0.056&&-0.045$\pm$0.021& 0.002$\pm$0.065& 0.052$\pm$0.091&-0.081$\pm$0.028\\
108708& 0.023$\pm$0.031&-0.040$\pm$0.079& 0.006$\pm$0.040& 0.037$\pm$0.024& 0.026$\pm$0.035&-0.007$\pm$0.025\\
108996& 0.046$\pm$0.112&-0.007$\pm$0.054&-0.024$\pm$0.029& 0.018$\pm$0.016& 0.032$\pm$0.024& 0.000$\pm$0.010\\
109931& 0.009$\pm$0.015&-0.037$\pm$0.106& 0.026$\pm$0.031& 0.014$\pm$0.021& 0.017$\pm$0.064& 0.059$\pm$0.030\\
113357&-0.015$\pm$0.026& 0.005$\pm$0.031&-0.016$\pm$0.022& 0.009$\pm$0.021&-0.011$\pm$0.037& 0.021$\pm$0.013\\
115604&-0.006$\pm$0.035& 0.003$\pm$0.039& 0.036$\pm$0.015& 0.031$\pm$0.015& 0.016$\pm$0.018& 0.031$\pm$0.039\\
118159& 0.006$\pm$0.033& 0.070$\pm$0.053& 0.000$\pm$0.055& 0.015$\pm$0.026&-0.007$\pm$0.019&-0.073$\pm$0.029\\

\hline
\end{tabular}
\label{t:abundances2}
\end{table*}
}

\onltab{4}{
\begin{table*}
\centering \tiny
\caption{Abundance ratios [X/Fe] for Ni, Cu, Zn, Y, Zr, and Ba. Errors correspond to the 1-$\sigma$ scatter of the line-by-line abundances. A conservative value of 0.050 for the error was assumed when only one line was available.}
\begin{tabular}{rrrrrrr} \hline\hline
HIP & \multicolumn{1}{c}{Ni} & \multicolumn{1}{c}{Cu} & \multicolumn{1}{c}{Zn} & \multicolumn{1}{c}{Y} & \multicolumn{1}{c}{Ba} & \multicolumn{1}{c}{S} \\ \hline
   348&-0.001$\pm$0.047&-0.003$\pm$0.030& 0.046$\pm$0.050&-0.046$\pm$0.041&-0.096$\pm$0.050&-0.005$\pm$0.014\\
   996& 0.019$\pm$0.028& 0.056$\pm$0.062& 0.116$\pm$0.050&-0.039$\pm$0.025&-0.046$\pm$0.050&-0.013$\pm$0.021\\
  1499& 0.057$\pm$0.060& 0.016$\pm$0.069& 0.027$\pm$0.050&-0.034$\pm$0.036&-0.086$\pm$0.050&-0.013$\pm$0.047\\
  2131&-0.030$\pm$0.023&-0.027$\pm$0.025& 0.095$\pm$0.050&-0.121$\pm$0.027&-0.130$\pm$0.050&-0.025$\pm$0.025\\
  2894&-0.022$\pm$0.036&-0.049$\pm$0.054&-0.008$\pm$0.050& 0.078$\pm$0.025& 0.002$\pm$0.050& 0.136$\pm$0.025\\
  4909&-0.052$\pm$0.039&-0.070$\pm$0.046&-0.065$\pm$0.050& 0.055$\pm$0.016& 0.008$\pm$0.050& 0.204$\pm$0.044\\
  5134&-0.041$\pm$0.033&-0.066$\pm$0.013&-0.026$\pm$0.050& 0.061$\pm$0.051& 0.081$\pm$0.050& 0.134$\pm$0.054\\
  6407&-0.023$\pm$0.039&-0.035$\pm$0.081&-0.015$\pm$0.050& 0.045$\pm$0.025& 0.076$\pm$0.050& 0.143$\pm$0.016\\
  7245&-0.011$\pm$0.043&-0.034$\pm$0.022& 0.027$\pm$0.050& 0.080$\pm$0.025& 0.065$\pm$0.050& 0.113$\pm$0.057\\
  8507&-0.034$\pm$0.025&-0.073$\pm$0.025&-0.046$\pm$0.050& 0.078$\pm$0.021&& 0.110$\pm$0.021\\
  8841&-0.028$\pm$0.030&-0.072$\pm$0.011& 0.072$\pm$0.050&-0.071$\pm$0.016& 0.071$\pm$0.050& 0.004$\pm$0.025\\
  9349&-0.037$\pm$0.029&-0.069$\pm$0.030&-0.042$\pm$0.050& 0.078$\pm$0.018& 0.054$\pm$0.050& 0.151$\pm$0.028\\
 10710&-0.012$\pm$0.053&-0.070$\pm$0.016& 0.006$\pm$0.050& 0.022$\pm$0.042& 0.037$\pm$0.050& 0.083$\pm$0.025\\
 11728&-0.012$\pm$0.032&-0.035$\pm$0.061& 0.011$\pm$0.050&-0.011$\pm$0.025& 0.005$\pm$0.050& 0.047$\pm$0.018\\
 11915&-0.026$\pm$0.038&-0.064$\pm$0.041&-0.071$\pm$0.050& 0.004$\pm$0.025& 0.130$\pm$0.050& 0.110$\pm$0.011\\
 14614&-0.022$\pm$0.042&-0.050$\pm$0.025& 0.058$\pm$0.050& 0.009$\pm$0.040& 0.070$\pm$0.050& 0.084$\pm$0.025\\
 14632& 0.030$\pm$0.039& 0.066$\pm$0.025& 0.042$\pm$0.050& 0.009$\pm$0.025&-0.111$\pm$0.050& 0.003$\pm$0.016\\
 18261&-0.032$\pm$0.022&-0.051$\pm$0.057& 0.057$\pm$0.050& 0.099$\pm$0.022& 0.064$\pm$0.050& 0.133$\pm$0.011\\
 22528&-0.018$\pm$0.049& 0.022$\pm$0.045& 0.215$\pm$0.050&-0.095$\pm$0.066& 0.048$\pm$0.050&-0.041$\pm$0.025\\
 23835&-0.013$\pm$0.029& 0.047$\pm$0.054& 0.088$\pm$0.050& 0.094$\pm$0.036& 0.123$\pm$0.050& 0.030$\pm$0.035\\
 25670&-0.016$\pm$0.027&-0.047$\pm$0.027&-0.007$\pm$0.050& 0.001$\pm$0.011& 0.016$\pm$0.050& 0.049$\pm$0.018\\
 28336&-0.024$\pm$0.042&-0.065$\pm$0.025&-0.091$\pm$0.050&-0.008$\pm$0.025& 0.072$\pm$0.050& 0.072$\pm$0.018\\
 38072&-0.005$\pm$0.032&-0.014$\pm$0.079&-0.098$\pm$0.050& 0.036$\pm$0.025& 0.064$\pm$0.050& 0.114$\pm$0.023\\
 38228&-0.048$\pm$0.099& 0.709$\pm$0.050&-0.147$\pm$0.050& 0.101$\pm$0.045& 0.018$\pm$0.050& 0.303$\pm$0.025\\
 42438&-0.093$\pm$0.084& 0.772$\pm$0.050&-0.166$\pm$0.050& 0.039$\pm$0.050&-0.049$\pm$0.050& 0.289$\pm$0.045\\
 44324&-0.027$\pm$0.090&-0.115$\pm$0.035&-0.026$\pm$0.050& 0.124$\pm$0.052& 0.030$\pm$0.050& 0.166$\pm$0.050\\
 46066&-0.045$\pm$0.029&-0.042$\pm$0.047& 0.003$\pm$0.050& 0.070$\pm$0.028& 0.082$\pm$0.050& 0.117$\pm$0.012\\
 49572&-0.007$\pm$0.033&-0.011$\pm$0.057& 0.065$\pm$0.050&-0.058$\pm$0.047&-0.005$\pm$0.050&-0.012$\pm$0.011\\
 49756&-0.010$\pm$0.050&-0.046$\pm$0.025&-0.088$\pm$0.050& 0.064$\pm$0.016&-0.020$\pm$0.050& 0.041$\pm$0.035\\
 52040&-0.030$\pm$0.025&-0.065$\pm$0.025&-0.001$\pm$0.050& 0.103$\pm$0.064& 0.025$\pm$0.050& 0.156$\pm$0.025\\
 52137&-0.049$\pm$0.024&-0.109$\pm$0.025&-0.105$\pm$0.050& 0.135$\pm$0.025& 0.087$\pm$0.050& 0.207$\pm$0.011\\
 55459&-0.008$\pm$0.016&-0.024$\pm$0.018&-0.019$\pm$0.050&-0.000$\pm$0.032&-0.031$\pm$0.050& 0.013$\pm$0.025\\
 56948&-0.006$\pm$0.036&-0.032$\pm$0.025&-0.015$\pm$0.050& 0.027$\pm$0.016& 0.110$\pm$0.050&-0.010$\pm$0.033\\
 56997&-0.063$\pm$0.017&-0.116$\pm$0.023&-0.078$\pm$0.050& 0.094$\pm$0.027& 0.089$\pm$0.050& 0.174$\pm$0.035\\
 60314&-0.041$\pm$0.054&-0.069$\pm$0.016&-0.152$\pm$0.050& 0.093$\pm$0.023& 0.037$\pm$0.050& 0.039$\pm$0.027\\
 62175&-0.027$\pm$0.020&-0.033$\pm$0.011&-0.004$\pm$0.050& 0.080$\pm$0.039&-0.002$\pm$0.050& 0.085$\pm$0.052\\
 64150&-0.019$\pm$0.025& 0.033$\pm$0.016&-0.048$\pm$0.050&& 0.148$\pm$0.050& 0.147$\pm$0.025\\
 64497&-0.057$\pm$0.071& 0.736$\pm$0.050&-0.144$\pm$0.050& 0.060$\pm$0.013&& 0.214$\pm$0.063\\
 64794&-0.028$\pm$0.027&-0.019$\pm$0.017& 0.101$\pm$0.050&-0.112$\pm$0.025& 0.016$\pm$0.050&-0.051$\pm$0.048\\
 72659&-0.101$\pm$0.035&-0.120$\pm$0.059&-0.065$\pm$0.050& 0.091$\pm$0.056& 0.025$\pm$0.050& 0.301$\pm$0.025\\
 73815&-0.012$\pm$0.022& 0.001$\pm$0.024& 0.046$\pm$0.050&-0.085$\pm$0.023&-0.084$\pm$0.050&-0.014$\pm$0.035\\
 74341&-0.014$\pm$0.025&-0.043$\pm$0.025&-0.123$\pm$0.050& 0.036$\pm$0.031& 0.076$\pm$0.050& 0.054$\pm$0.025\\
 77466& 0.019$\pm$0.042& 0.042$\pm$0.019& 0.227$\pm$0.050&-0.003$\pm$0.024&-0.021$\pm$0.050&-0.075$\pm$0.014\\
 78028&-0.000$\pm$0.062& 0.011$\pm$0.045&-0.012$\pm$0.050& 0.096$\pm$0.025& 0.165$\pm$0.050& 0.019$\pm$0.059\\
 78680&-0.089$\pm$0.040& 0.647$\pm$0.050&-0.039$\pm$0.050& 0.118$\pm$0.011& 0.141$\pm$0.050& 0.222$\pm$0.015\\
 79186&-0.034$\pm$0.040& 0.026$\pm$0.025& 0.103$\pm$0.050&-0.039$\pm$0.023&-0.144$\pm$0.050&-0.027$\pm$0.011\\
 79672&-0.026$\pm$0.011&-0.022$\pm$0.049& 0.012$\pm$0.050& 0.060$\pm$0.022& 0.121$\pm$0.050& 0.068$\pm$0.025\\
 81512&-0.008$\pm$0.036& 0.036$\pm$0.028& 0.058$\pm$0.050&-0.025$\pm$0.036& 0.039$\pm$0.050& 0.032$\pm$0.014\\
 83601&-0.041$\pm$0.063& 0.731$\pm$0.050&-0.164$\pm$0.050& 0.058$\pm$0.025& 0.010$\pm$0.050& 0.154$\pm$0.058\\
 88194&-0.022$\pm$0.021& 0.003$\pm$0.011&-0.058$\pm$0.050&-0.028$\pm$0.018&-0.065$\pm$0.050& 0.024$\pm$0.014\\
 88427& 0.013$\pm$0.022& 0.028$\pm$0.014& 0.141$\pm$0.050&-0.044$\pm$0.042& 0.057$\pm$0.050&-0.032$\pm$0.033\\
 89443&-0.017$\pm$0.030&-0.004$\pm$0.025&-0.020$\pm$0.050&-0.017$\pm$0.021& 0.048$\pm$0.050&-0.007$\pm$0.013\\
 96895& 0.013$\pm$0.018& 0.058$\pm$0.021& 0.032$\pm$0.050&-0.018$\pm$0.011&-0.030$\pm$0.050&-0.013$\pm$0.025\\
 96901& 0.014$\pm$0.021& 0.037$\pm$0.013&-0.004$\pm$0.050&-0.058$\pm$0.014&-0.019$\pm$0.050&-0.041$\pm$0.020\\
100963&-0.029$\pm$0.028&-0.057$\pm$0.025&-0.046$\pm$0.050& 0.100$\pm$0.030& 0.050$\pm$0.050& 0.108$\pm$0.025\\
102152&-0.013$\pm$0.027& 0.018$\pm$0.025& 0.073$\pm$0.050&-0.081$\pm$0.025&-0.099$\pm$0.050&-0.017$\pm$0.025\\
104504&-0.066$\pm$0.044&-0.094$\pm$0.065&-0.019$\pm$0.050& 0.088$\pm$0.012& 0.051$\pm$0.050& 0.269$\pm$0.025\\
107350&-0.090$\pm$0.084&&-0.142$\pm$0.050& 0.055$\pm$0.050&& 0.221$\pm$0.035\\
108708&-0.052$\pm$0.029&-0.107$\pm$0.040&-0.097$\pm$0.050& 0.025$\pm$0.043& 0.031$\pm$0.050& 0.175$\pm$0.054\\
108996&-0.035$\pm$0.053&-0.089$\pm$0.035&-0.122$\pm$0.050& 0.064$\pm$0.037& 0.127$\pm$0.050& 0.150$\pm$0.057\\
109931& 0.026$\pm$0.039& 0.039$\pm$0.025& 0.033$\pm$0.050&-0.065$\pm$0.025&-0.093$\pm$0.050&-0.035$\pm$0.045\\
113357& 0.021$\pm$0.047& 0.027$\pm$0.055& 0.004$\pm$0.050&-0.027$\pm$0.011& 0.022$\pm$0.050&-0.045$\pm$0.025\\
115604& 0.031$\pm$0.024& 0.057$\pm$0.038& 0.069$\pm$0.050&-0.050$\pm$0.033&-0.049$\pm$0.050&-0.034$\pm$0.011\\
118159&-0.052$\pm$0.053&-0.124$\pm$0.049&-0.131$\pm$0.050& 0.127$\pm$0.025& 0.137$\pm$0.050& 0.251$\pm$0.014\\

\hline
\end{tabular}
\label{t:abundances3}
\end{table*}
}

\section{Abundance trends among solar twin stars}

In Fig.~\ref{f:gce} (available online), we show the abundance ratios [X/Fe] determined in this work as a function of $\feh$. When compared to the abundance trends for Galactic thin-disk stars published elsewhere \citep[e.g.,][]{bensby05,reddy06,takeda07:abundances,neves09}, we notice that the slopes and relative scatter (between different elements) are compatible with our results (all but the two most metal-poor stars in our sample have thin-disk kinematics). However, there is no general agreement about the zero points of the abundance ratios, which is a consequence of systematic errors that affect the various techniques used by other authors differently. Since those errors have been minimized in our differential work, the zero points of the abundance scales reported here should be reliable.

\onlfig{1}{
\begin{figure*}
\includegraphics[bb=70 380 670 870,width=15.8cm]{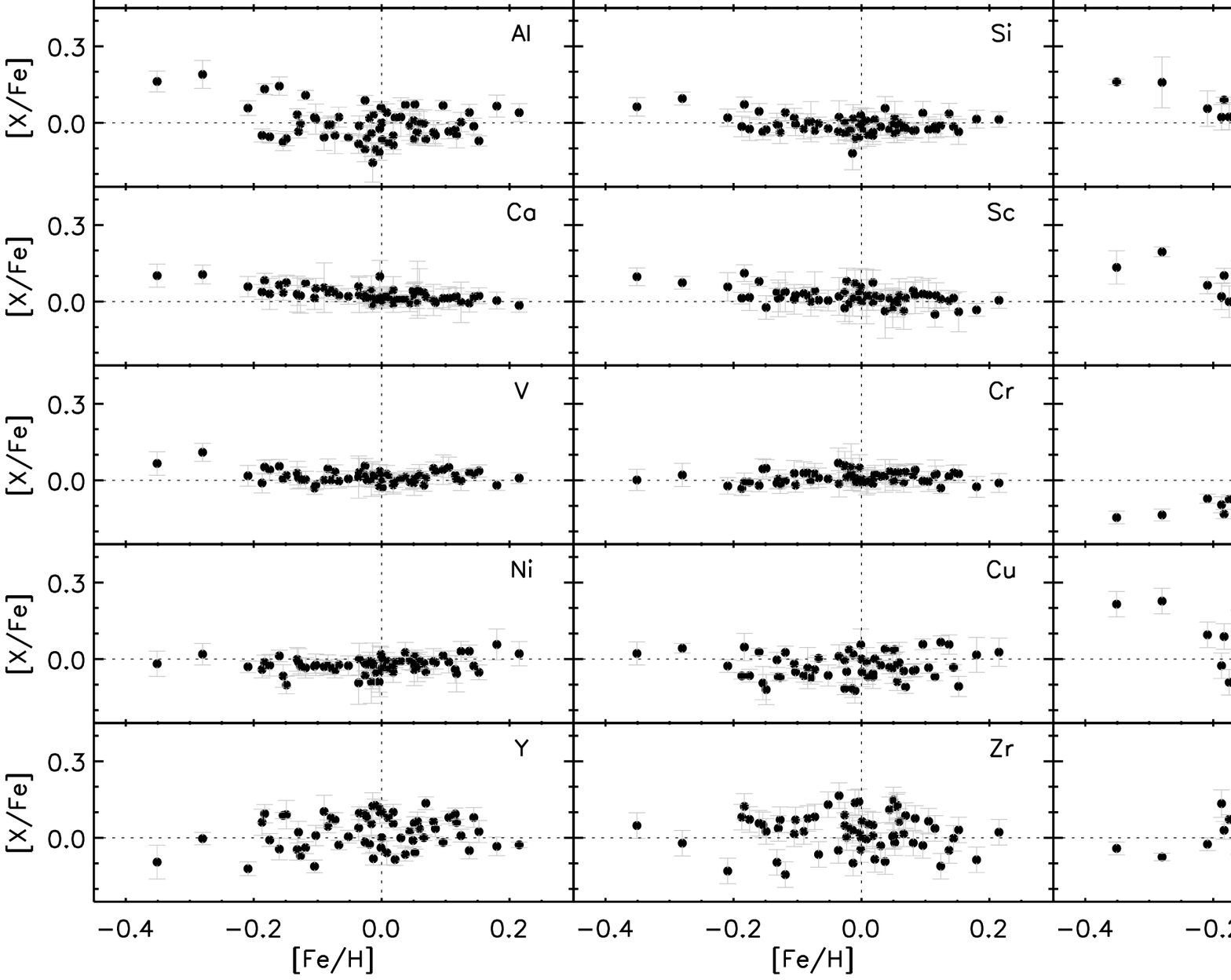}
\caption{Abundance ratios as a function of iron abundance for full sample of 64 stars. Dotted lines denote solar values.}
\label{f:gce}
\end{figure*}
}

The decreasing [X/Fe] vs. $\feh$ trends for C, O, S, Ca, Sc, and Ti are consistent with currently accepted interpretations of thin-disk chemical evolution \citep[e.g.,][]{mcwilliam97}. Al, Si, and Zn also show this trend but the slope is too shallow to produce a noticeable effect in our data, given the short $\feh$ range. This is attributed to the increasing importance of the Type Ia supernovae (SNe) contribution to the interstellar medium (ISM) composition compared to Type II SNe. The very steep increase in [Mn/Fe] with $\feh$ can be understood as metallicity-dependent yields from SNe II. The nearly constant Na, V, Cr, Ni, and Cu abundance ratios stem from these elements and Fe roughly having the same nucleosynthetic origins. The abundance ratios of the $s$-process elements Y, Zr, and in particular Ba, which are thought to be produced mainly in AGB stars, are strongly dependent on stellar age \citep[e.g.,][]{edvardsson93,bensby07}; therefore, the large scatter seen is likely related to the age span of our sample. Also, non-negligible non-LTE effects are predicted for Ba \citep[$\simeq0.1$\,dex,][]{mashonkina99,mashonkina00}, but they have not been corrected here.

\begin{figure}
\includegraphics[bb=75 377 340 742,width=9.3cm]{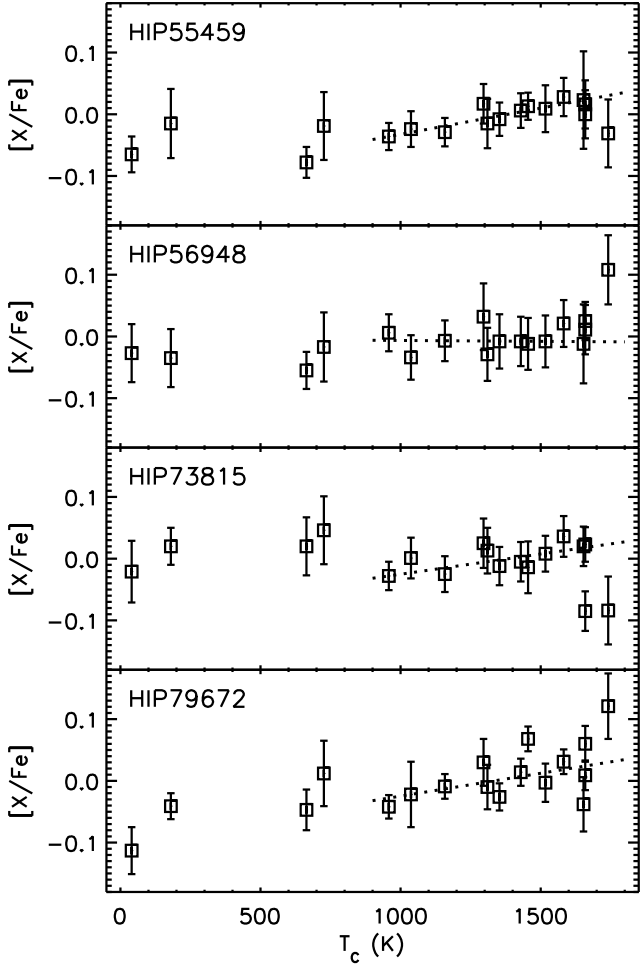}
\caption{Abundance ratios as a function of condensation temperature ($\tc$) for four of our solar twins. The dotted line is a linear fit to the abundance ratios of refractory elements ($\tc>900$\,K).}
\label{f:trend_stars}
\end{figure}

\begin{figure}
\includegraphics[bb=65 370 450 610,width=9.3cm]{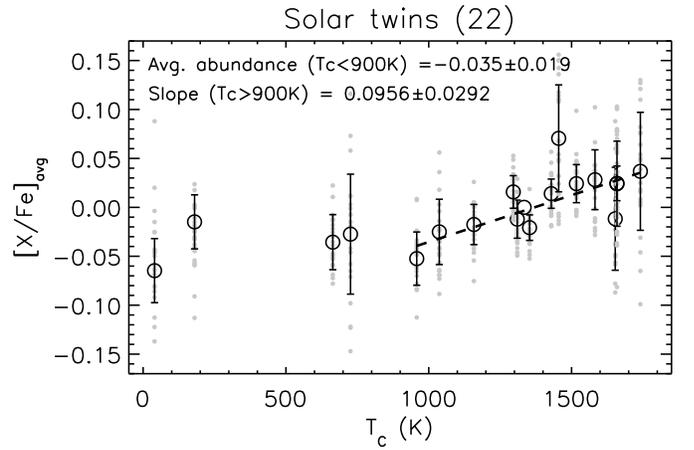}
\caption{Average abundance ratios of 22 solar twin stars as a function of condensation temperature ($\tc$). Gray solid circles represent individual abundances, while open circles with error bars correspond to the weighted average and standard deviation of all stars for each element. The open circle without error bar corresponds to iron. The dashed line is a linear fit to the abundance ratios of refractory elements.}
\label{f:trend_sun_avg}
\end{figure}

Next, we examined the relation between [X/Fe] and condensation temperature ($\tc$; calculated for a solar-system composition gas by \citealt{lodders03}). Most of our sample stars show a significant correlation, although individually the scatter is relatively large (Fig.\,~\ref{f:trend_stars}). Thus, we averaged the [X/Fe] values for our 22 solar twins and compared them to $\tc$. As seen in Fig.~\ref{f:trend_sun_avg}, refractory elements ($\tc>900$\,K) are overabundant with respect to the volatiles ($\tc<900$\,K) in the solar twins compared to the Sun by as much as 20\,\% ($\simeq0.08$\,dex). A linear fit to the $\tc>900$\,K elements, weighted by the star-to-star scatter, shows that the abundance vs. $\tc$ correlation of refractory elements is significant at the 3\,$\sigma$ level. The elements that depart the most from this trend are Al ($\tc=1653$\,K) and Ba ($\tc=1455$\,K). For Ba, a combination of age-related and strong non-LTE effects could be responsible for the large star-by-star scatter seen in Fig.~\ref{f:gce} \citep[e.g.,][]{bensby07,mashonkina00}.

Following the referee's suggestion, we also examined the relation between [X/Fe] and the first ionization potential (FIP). While at first sight a correlation is apparent, its significance is less than for the $\tc$ trend ($1.5\,\sigma$ for the $\tc>900$\,K elements). We note, in particular, that Na, an element that clearly defines the $\tc$ trend, stands out as an obvious outlier on the FIP correlation. Moreover, if the abundances are plotted against the ionization potential of the species that correspond to the spectral lines used in our work (Y, Zr, and Ba abundances were derived from their singly ionized lines), the (already weak) correlation disappears.

\begin{figure}
\includegraphics[bb=50 365 340 585,width=9.0cm]{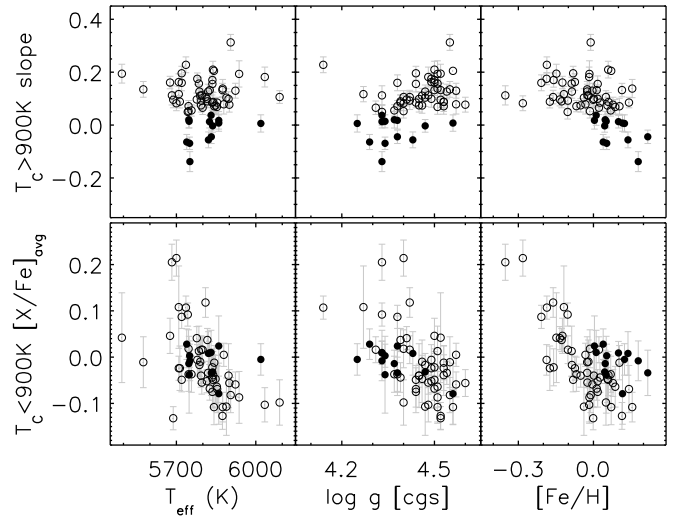}
\caption{Top panel: abundance vs. $\tc$ slope for $\tc>900$\,K as a function of stellar parameters. Bottom panel: average abundance of $\tc<900$\,K elements as a function of stellar parameters. The filled circles are stars that have $\tc$ slopes less than their 1-$\sigma$ error, including stars with negative slope. Units of $\tc$ slope are $10^{-3}$\,dex\,K$^{-1}$.}
\label{f:abund_stats}
\end{figure}

To further examine the correlation with $\tc$, we show the relation between the abundance ratio vs.~$\tc$ slope for the refractory elements ($\tc>900$\,K) and stellar parameters in Fig.~\ref{f:abund_stats}. We also show there the relation between the average abundance ratio of volatiles ($\tc<900$\,K) and stellar parameters. The $\tc>900$\,K slopes and $\tc<900$\,K average abundances plotted in Fig.~\ref{f:abund_stats} are listed in Table~\ref{t:parameters}, available online. The trend seen for the average abundance of volatiles vs. $\feh$ relation comes from chemical evolution effects, given that the abundance ratios of C and O increase with lower $\feh$ (Fig.~\ref{f:gce}). The $\tc>900$\,K slope is correlated with the stellar surface gravity ($\logg$) and $\feh$. Furthermore, at super-solar metallicities, there seem to be two distinct groups of stars, one showing positive slope and underabundance of volatiles and the other one having negative slope and roughly solar abundance of volatiles. A Kolmogorov-Smirnov test shows that the likelihood that the distribution of slopes at $\feh>0$ corresponds to a single Gaussian centered around zero is about 2.7 times less than that of two Gaussians centered on $-0.05$ and $0.09\times10^{-3}$\,dex\,K$^{-1}$, respectively (assuming a width of $0.03\times10^{-3}$\,dex\,K$^{-1}$, which is the average observed value of our 22 solar twins).

Given the very limited $\feh$ range ($=0.0\pm0.1$\,dex), Galactic chemical evolution effects are unimportant for our solar twins sample, and so the trend with $\tc$ shown in Fig.~\ref{f:trend_sun_avg} is most likely related to other processes. Similar abundance trends with $\tc$ have been found for the ISM \citep[e.g.,][]{savage96} and for certain types of objects such as post-AGB, RV Tauri, and $\lambda$~Bootis stars \citep[e.g.,][]{venn90,giridhar05}, albeit with much larger amplitudes. In the case of $\lambda$~Bootis stars, the $\tc$ trend is attributed to the accretion of volatile rich gas with refractories depleted into grains in a fashion qualitatively consistent with ISM depletion patterns. In the atmospheres of the majority of our solar twins, elements that are more likely to form dust grains are more overabundant compared to the Sun. If the Sun and its twins formed similarly (in particular from gas with very similar initial chemical composition), where did the refractory elements go in the solar case?

\section{Planetary signatures in the abundance trends?}

\cite{melendez09:twins}, who were the first to detect the $\tc$ abundance trends that we confirm in this paper, suggest that the solar chemical composition has been affected by the formation of planets. Inner solar system objects are enriched in refractory elements relative to volatiles \citep[e.g,][]{palme00}, mirroring the trend seen in Fig.~\ref{f:trend_sun_avg}. Relative abundances in meteorites are roughly constant up to $\tc\simeq1000$\,K but increase steeply with $\tc$ for $\tc\gtrsim1000$\,K \citep[][their Fig.~2]{alexander01}. In the solar twins compared to the Sun, the overabundance of refractory elements relative to volatiles is of about 20\,\% (or 0.08\,dex). An order of magnitude calculation shows that such a difference would roughly  disappear if the total mass of refractory elements in the terrestrial planets of the solar system today were to be added to the solar convective zone \citep[][]{melendez09:twins}. If the majority of solar twins did not form planets, their original chemical compositions should not have been altered, with no deficiency of refractories, as observed. The near constancy of the abundance of volatiles ($\tc\lesssim1000$\,K) and the fact that they do not condense at 
the high temperatures present in the inner proto-solar system suggest that the Sun and its twins have retained those elements. The abundance differences for the volatiles would dissappear on average had we chosen one of those elements as the reference element rather than the refractory element Fe.

The behavior of the $\tc>900$\,K slope with metallicity is particularly interesting (Fig.~\ref{f:abund_stats}). At solar and sub solar metallicity, few stars have near zero slope (within their 1-$\sigma$ error). At super-solar metallicities, however, there seem to be two groups of stars: one showing positive $\tc$ slope and the other one negative. Following our line of reasoning, a negative slope implies that an even greater fraction of refractory elements have been extracted from the star-forming cloud to make up dust grains, also suggesting planet formation. Since the frequency of planetary systems increases with $\feh$ \citep[e.g.,][]{udry07}, it is tempting to conclude that this bimodal distribution is separating super-solar metallicity stars with and without terrestrial planets. Thus, while for $\feh\lesssim0.1$ the fraction of stars showing the proposed ``planet signature'' in their chemical composition is $\simeq15$\,\%, for $\feh>0.1$ the number is $\simeq50$\,\% or more. The low number of stars analyzed at high $\feh$ prevent us from determining this number more accurately.

Certainly, our interpretation of the $\tc$ trend may be questioned. For example, that the currently accepted values for the lifetimes of observed disks around pre-main-sequence stars are too short \citep[$\sim3$ to 10\,Myr; e.g.,][]{meyer09}, compared to the time it took the solar convective zone to reach its present size \citep[$\sim30$\,Myr; e.g.,][]{dantona94}, suggests that the planet signature could not have been imprinted into the photospheric composition. However, recent realistic star formation calculations suggest that the structure of the early Sun was similar to that of the present one, and never fully convective \citep{wuchterl01,wuchterl03}, which would solve the time-scale problem \citep{nordlund09}.

Independently of the interpretation, the observational result that the solar chemical composition is anomalous when compared to solar-type stars is very robust. We have minimized systematic errors by using differential analysis and reduced observational scatter by averaging the results over many stars that are similar to each other.

Previous studies \citep[e.g.,][]{smith01,takeda01,santos04,ecuvillon06,gonzalez07} have been cautious about reaching strong conclusions based on abundance trends with $\tc$ because of the still relatively large systematic errors associated with their analyses in comparison with the small size of the effect. To detect this trend, a precision of $\simeq0.03$\,dex or better is required. In fact, as stated above, the $\tc$ trend shown in Fig.~\ref{f:trend_sun_avg} was first detected by \cite{melendez09:twins}, who were able to achieve a precision of 0.01\,dex in their derived abundance ratios using spectra of higher quality for eleven southern hemisphere solar twins.

Detecting planets around other stars is one of the major challenges of contemporary astrophysics. Current technology allows us to find giant planets in close-in orbits with relative ease, so remarkable progress is being made in our understanding of exoplanets \citep[e.g.,][]{udry07}. Terrestrial planets and solar system analogs, however, remain elusive. The possibility of identifying them using detailed chemical composition analyses is therefore very promising. There is no reason to restrict this experiment to solar twins, because any group of stars twins of each other can be used to measure very precise relative abundances. We are currently working on the homogeneous (i.e., same telescope/instrument/observing conditions) acquisition and analysis of very high-quality spectra ($R\simeq100,000$; $S/N\gtrsim400$) of stars with and without detected planets to continue this investigation.


\bibliographystyle{aa}


\end{document}